\documentclass[aip,jap,preprint]{revtex4}

\usepackage{graphicx}
\usepackage{dcolumn}
\usepackage{bm}
\usepackage{amsmath}
\usepackage{amssymb}
\usepackage{latexsym}
\usepackage{epsfig}
\usepackage{amsbsy}
\usepackage{array}
\usepackage{amssymb}
\usepackage{setspace}
\usepackage{bm}
\usepackage{subfigure}

\def\sint{\ifmmode{- \!\!\!\!\!\! \int}
    \else{\hbox{$- \!\!\!\! \int \ $}}\fi}

\begin{document}

\title
{ Thermodynamic properties of LiCu$_{2}$O$_{2}$ multiferroic
compound}

\author{Yan Qi}
\author{An Du\footnote{Author to whom correspondence should be addressed.
Electronic mail: duanneu@163.com. Tel.: 086-024-83678326.}}
\affiliation{%
College of Sciences, Northeastern University, Shenyang 110819, China
}%


\begin{abstract}
A spin model of quasi-one dimensional LiCu$_{2}$O$_{2}$ compound
with ground state of ellipsoidal helical structure in which the
helical axis is along the diagonal of CuO$_{4}$ squares has been
adopted. By taking into account the interchain coupling and exchange
anisotropy, the exotic magnetic properties and ferroelectricity
induced by spiral spin order have been studied by performing Monte
Carlo simulation. The simulation results qualitatively reproduce the
main characters of ferroelectric and magnetic behaviors of
LiCu$_{2}$O$_{2}$ compound and confirm the low-temperature
incollinear spiral ordering. Furthermore, by performing the
calculations of spin structure factor, we systematically investigate
the effects of different exchange coupling on the lower-temperature
magnetic transition, and find that the spiral spin order depends not
only on the ratio of nearest and next-nearest neighbor inchain spin
coupling but also strongly on the exchange anisotropy.
\end{abstract}

\maketitle

\section{Introduction}

The multiferroics, in which magnetism and ferroelectricity coexist
in the same material, have attracted many researchers since 1960s
due to their fundamental physics and potential technological
applications~\cite{tk,nh}. In recent years, the magnetism-driven
ferroelectricity discovered in frustrated magnets renews the
interest in this field. A simple prototype of frustrated magnets is
a spin chain with the nearest-neighbor (NN) ferromagnetic (FM)
coupling and the next-nearest-neighbor (NNN) antiferromagnetic (AF)
coupling. Such competing interactions can lead to frustration and an
incommensurate spiral order in magnetic materials~\cite{rbga}.
Actually, a class of low-dimensional cuprate oxides such as
LiCuVO$_{4}$, Li$_{2}$CuO$_{2}$, Li$_{2}$ZrCuO$_{4}$ etc. were
reported to have such spiral magnetic
orders~\cite{me,sl,mg,sp,aa,slj}. And it is believed that the
research on the ferroelectricity of magnetic origin in these
low-dimensional compounds will be helpful in understanding the
multiferroic nature~\cite{yy,yk}.

LiCu$_{2}$O$_{2}$ is a typical representative of multiferroic
cuprates with a quantum spin $S$=1/2~\cite{kyc,hch}. It has an
orthorhombic crystal structure with a \textit{Pmna} space group and
unit cell parameters $a$$=$5.73 {\AA}, $b$$=$2.86 {\AA}, $c$$=$12.42
{\AA}. The crystal structure is twinned at the microscopic level in
the $ab$ plane as a result of the unit cell parameter $a$ very close
to $2b$~\cite{lz}. It has an equal number of Cu$^{2+}$ and Cu$^{1+}$
ions in distinct nonequivalent crystallographic positions. The
magnetic Cu$^{2+}$ ions are located at the center of edge-sharing
CuO$_{4}$ plaquettes and form a zigzag like spin-chain structure
along the $b$ axis~\cite{tma,tm,lmb}. The experimental data reveals
that the system undergose two successive magnetic transitions at
$T_{N1}$$\sim$25K and $T_{N2}$$\sim$23K~\cite{ari,ssy}. Below
$T_{N2}$, the ferroelectricity emerges with the spiral magnetic
order. At $T$$_{N2}$$<$T$<$$T$$_{N1}$, the intermediate state is
found to be a sinusoidal spin state.

Although many experiments and theoretical studies on
LiCu$_{2}$O$_{2}$ have been performed~\cite{dvd,mhq,jsi,mc,rma}, the
nature of the ground state and the origin of the ferroelectricity
remain under debate. Masuda et al. carried out an unpolarized
neutron diffraction study and firstly proposed the incommensurate
helimagnetic order in $ab$ plane~\cite{tma}. However, based on this
$ab$-spiral picture, the fact that the polarization emerges along
the $c$ axis can not be interpreted by the commonly accepted
microscopic mechanisms: the inverse Dzyaloshinskii-Moriya
interaction or spin-current model. Later, Park et al. suggested
another spin picture that the transverse spiral spin component was
in the $bc$-plane~\cite{sp}, which is partially confirmed by Seki et
al., but the quantitative calculation on the intensity of polarized
neutron reflections shows a prominent discrepancy~\cite{ssy}.
Recently, based on their own NMR and neutron diffraction data, Sato
et al. proposed a new noncollinear modulated magnetic structure.
They pointed out that the magnetic structure of LiCu$_{2}$O$_{2}$ in
the low-temperature phase ($T$$<$$T_{N2}$) is an ellipsoidal
structure with the helical axis tilted by about 45$^{\circ}$ from
the $a$- or $b$- axis within the $ab$ plane~\cite{yy,yk}. This
45$^{\circ}$-tilt spin model receives strong evidence support from
the very recent experiments performed by Li et al~\cite{lz}.
However, as far as we know, the corresponding theoretical
investigations based on this new model are still absent.

In this paper, with the consideration of interchain coupling and
exchange anisotropy, a spin model of LiCu$_{2}$O$_{2}$ with ground
state of the ellipsoidal helical structure has been employed. Monte
Carlo simulation is carried out to investigate the fascinating
magnetoelectric coupling behaviors in this multiferroic compound.
Our simulation qualitatively reproduces the experimental results of
the complicated electric and magnetic behaviors observed in
LiCu$_{2}$O$_{2}$, confirming the spiral spin order at ground state.
In addition, the influences of the different exchange interactions
and easy-plane anisotropies on the lower-temperature magnetic phase
pattern have also been explored. The spin structure factors are
calculated to analyze the variations of the underlying microscopic
magnetic and ferroelectric structures. We believe that the present
work will be helpful in understanding the ferroelectricity in
frustrated magnets and shed some additional light on this
interesting physical subject.

\section{Model and simulation}
LiCu$_{2}$O$_{2}$ is a complex spin-driven ferroelectricity
multiferroic compound. Its magnetism stems from the two linear
Cu$^{2+}$ chains, which propagate along the $b$ axis and form a
zigzag ladderlike structure. These ladders are isolated from each
other by both Li ions and the layers of non-magnetic Cu$^{1+}$ ions.
The sketch map of the magnetic structure is demonstrated in Fig. 1,
which can be regarded as an equivalent quasi-one dimensional Bravais
lattice of spins~\cite{tma,tm}. The Hamiltonian of this system can
be written in the following general form \begin{equation}
H=\sum_{i,j}J_{ij}(S_{i}^{a}S_{j}^{a} +S_{i}^{b}S_{j}^{b}+\Delta
S_{i}^{c}S_{j}^{c})-D\sum_{i}(\bm{S}_{i}\cdot
\bm{e}_{i})^{2}-\bm{h}\cdot\sum_{i} \bm{S}_{i}-\bm{h_{e}}\cdot
\sum_{i}\bm{P_{i}},\label{haml} \end{equation} where
($S_{i}^{a}$,$S_{i}^{b}$,$S_{i}^{c}$) are the classical spin
components with unit vectors. $\Delta\leq 1$ is the exchange
anisotropy, usually presenting in low-dimensional system with
multiple magnetic transitions~\cite{xyc}. Theoretically, an XXZ-type
anisotropy ($\Delta$$\lesssim$1) is expected to stabilize a vector
chiral order~\cite{jsi}. And recent experiment also shows that
substantial exchange anisotropies exist in the edge-sharing
spin-chain compounds~\cite{hak}. In order to yield results that much
close to those of experiments, the value of $\Delta$=0.7 is adopted
here. $J_{ij}$ represents exchange coupling between spins on
different site. As a result of the Cu-O-Cu bond angles near
94$^{\circ}$, the NN FM inchain coupling $J_{2}$ and NNN AF inchain
coupling $J_{4}$ are expected from the Kanamori-Goodenough
rule~\cite{slj}. The estimated ratio $|J_{4}/J_{2}|$ has been
experimentally verified to vary from 0.50 to 0.65~\cite{tma,rma}.
However, upon the values of $J_{1}$ and $J_{\bot}$, it is still a
subject of discussion at present. Based on inelastic neutron
scattering experiments and spin wave theory, Masuda et
al.\cite{tmaz} suggest a strong AF "rung" interaction $J_{1}$ and a
weak interchain coupling $J_{\perp}$, while Drechsler et al. obtain
a contrary conclusion through the analysis of electronic structure
and cluster calculations~\cite{slj}. Irrespective of this dispute,
we adopt the suggestion of Masuda et al. here. Our simulation also
qualitatively demonstrates a good agreement with the results of
experiments under the condition of a strong $J_{1}$ coupling. Thus,
unless particularly stated, the default values of $J_{1}$, $J_{2}$,
$J_{4}$ and $J\perp$ are set as 3.4, -6.0, 3.0 and 0.9 respectively
in this paper, which is slightly different from those determined
from the spin-wave spectra in the proportion of exchange constants.

Considering the 45$^{\circ}$-tilt spin model proposed by Sato et al.
(Fig. 2), a plane anisotropy has been added in the Hamiltonian. $D$
represents the magnitude of the magnetic anisotropy. $\bm{e}_{i}$ is
a unit vector and in the direction of [110], representing the
direction of magnetic anisotropy. Due to the large negative value of
anisotropy, $D$=$-5$, a spiral magnetic order forming in the [110]
plane can be expected. And this kind of anisotropy will lead to a
strong spin coupling along $c$ axis, and therefore plays an
essential role for the emergency of spontaneous polarization
~\cite{swh}. According to the observation of the low temperature
magnetic structure, Sato et al.\cite{yy,yk} have confirmed that the
relation $\bm{P}\propto$$\bm{Q}\times\bm{e}_{3}$ holds in
LiCu$_{2}$O$_{2}$ compound, where $\bm{P}$, $\bm{Q}$ and
$\bm{e}_{3}$ are the ferroelectric polarization, the modulation
vector and the helical axis of the ordered spins. This indicates
that the theories derived by phenomenological\cite{mmo} and
microscopic models~\cite{hkn,cjs,hjx} are also applied for
polarization of LiCu$_{2}$O$_{2}$. Thus, according to the spin
current model, or equivalently, the inverse Dzyaloshinskii-Moriya
interaction, $\bm{P}_{i}$ induced by the neighboring canting spins
($\bm{S}_{i}$ and $\bm{S}_{j}$) is expressed as follows:
\begin{equation}
\bm{P}_{i}=-A\bm{e}_{ij}\times(\bm{S}_{i}\times\bm{S}_{j}),
\end{equation}
where $\bm{e}_{ij}$ denotes the vector connecting the two sites of
$\bm{S}_{i}$ and $\bm{S}_{j}$, namely in the direction of magnetic
modulation vector along the $b$ axis. $A$ is a proportional constant
determined by the spin-exchange and the spin-orbit interactions as
well as the possible spin-lattice coupling term. It is assumed to be
unity here. Judging from this relation, the spontaneous polarization
along the $c$ axis observed in the experiment can be expected.
Consequently, the third and fourth items in Hamiltonian
Eq.~(\ref{haml}) correspond to magnetic and electric energies, where
$\bm{h}$ is the external magnetic field including an extra factor
$g\mu_{B}$ and $\bm{h}_{e}$ is the external electric field.

According to the statistical definitions and thermal fluctuations,
the general expressions for those quantities concerned in this paper
can be written in the following form:
\begin{eqnarray}
{m_{\alpha}}=\frac{1}{N}\sum_{i}S_{i}^{\alpha},\\
P_{\alpha}=\frac{1}{N}\sum_{i}P_{i}^{\alpha},\\
\chi_{m}^{\alpha}=\frac{1}{NT}(<M^{2}_{\alpha}>-<M_{\alpha}>^{2}),\\
\chi_{e}^{\alpha}=\frac{1}{NT}(<P^{2}_{\alpha}>-<P_{\alpha}>^{2}),\\
C_{m}=\frac{1}{NT^{2}}{(<H^{2}>-<H>^{2})}.
\end{eqnarray}
Here $\alpha$=$a,b,c$ labels the three axis respectively. $N$ is the
number of the particles. $m_{\alpha}$ and $P_{\alpha}$ denote the
average magnetization and polarization. $\chi_{m}^{\alpha}$ and
$\chi_{e}^{\alpha}$ are the average magnetic and electric
susceptibilities. $C_{m}$ is the average specific heat. $T$
represents the temperature and the Boltzmann constant $k_{B}$ is
absorbed into $T$.

We performed standard Monte Carlo simulation on an
$L$$\times$$L^{'}$ lattice with periodic boundary conditions.
$L$=100 is the length of a zigzag chain and $L^{'}$=10 is the number
of the zigzag chains. It is assumed that $a$, $b$ and $c$ axes are
respectively, along the directions of [100], [010], and [001]. The
spin is updated according to the Metropolis algorithm. For every T,
the initial 50000 Monte Carlo steps (MCS) are discarded for
equilibration, and then the results are obtained by averaging 15000
data. Each data is collected at every 10 MCS.

The final results are obtained by averaging twenty independent data
sets obtained by selecting different seeds for random number
generation.

\section{Results and discussions}
\subsection{Ground state configuration and size effects}

Similar to the measurement process in experiment~\cite{yk,yy}, the
system is initially polarized by a small electric field
${h}_{e}$=0.1 along the $c$ axis, which is implemented under the
condition of the zero magnetic field cooling (ZFC). After that, the
thermodynamic properties concerned in this paper are collected in a
warming process with ${h}_{e}=0.1$ along the $c$ axis. The purpose
of the poling procedure is to produce a single magnetoelectric
domain with magnetic modulation vector along the chain. In fact, the
spin-rotation axis of the spiral magnetic structure is with equal
probability along [110], [$\bar{1}$$\bar{1}$0], [1$\bar{1}$0], or
[$\bar{1}$10] direction. The choice depends on the sets of
parameters. In the present simulation, due to a large negative
magnetic anisotropy $D$=$-$5 in the diagonal of $ab$ plane, the
ground state composed of the spiral spin structure with spiral axis
along the [110] direction is generated.

Figure 3(a) presents the typical snapshot for the spin configuration
at lowest temperature, demonstrating the formation of the spiral
order at ground state. In order to analyze the spiral spin state
accurately, the zero-field spin structure factor $S(q)$ is evaluated
to scrutinize the microscopic magnetic structure and size effects.
Since this compound possesses a weak $J_{\perp}$ interaction and the
spiral order is formed along the chain direction, $L^{'}=10$ is
fixed in the present simulation for convenience. The lattice with
different sizes $L$=30$-$200 is examined by $S(q)$ which is
calculated along the chain direction with its expression written
as\cite{xy}
\begin{equation}
S(q)=\sum_{i,r}\cos({q}\cdot {r})\langle \bm{S}_{i}\cdot
\bm{S}_{i+r}\rangle,
\end{equation}
where $q$ is wave vector. $r$ is calculated in units of distance
between two nearest-neighbor correlated spins. In the inset of Fig.
3(b), $S(q)$ obtained in the lattice of $L=100$ at $T=0.01$ is
plotted. The sharp characteristic peaks appear at $q_{B}=0.3(\pi)$
and its equivalent position in $2\pi-q_{B}$, which confirm a good
spiral spin order formed in ground states, as shown in Fig. 3(a). In
addition, it is worth noted that the wavelength of spin structure
will be limited by the periodic boundary condition, therefore the
lattice size dependence of wave vector $q_{B}$ is examined to verify
the perfect spiral spin structure. As shown in Fig. 3(b), the value
of $q_{B}$ vibrates with $L$ varying, and the amplitude of vibration
decreases as $L$ increases. When $L$ exceeds 80, the wave vector
remains in the same position and the vibration disappears,
indicating the negligible size effects under this circumstance.
Thus, $L=100$ chose here is adequate for the discussion of
magnetoelectric properties in this system. To further confirm the
conclusion above, the thermal dependence of the bulk properties for
different lattice sizes under $h$$\parallel$$c$ are also displayed
in Fig. 4. It can be seen that the results obtained with lattice
size $L=100$ are almost identical with those obtained from larger
lattice size.
\subsection{Compare to experimental results}
In the following, we will concentrate on the magnetoelectric
properties of LiCu$_{2}$O$_{2}$ from two aspects: the macroscopic
thermodynamic behaviors and microscopic magnetic and ferroelectric
structures. For the former, the simulation results on the bulk
properties will be made a detail comparison with those of
experimental results to reveal the spiral spin order nature of the
ferroelectricity. As for the latter, our focus is on the influences
of the exchange couplings and anisotropies on the spiral spin
states, which will be explored by evaluating the spin structure
factors.

In Fig. 5, the temperature dependence of magnetization $m$ and
susceptibility $\chi_{m}$ under a small magnetic field are
presented. It can be seen that our simulation result is
qualitatively in good agreement with experimental data in two
aspects. First, $m$ ($\chi_{m}$) for $h\parallel c$ exceeds that for
$h\parallel a,b$, reproducing the anisotropic behaviors observed in
LiCu$_{2}$O$_{2}$~\cite{aa,yy}. Moreover, the negligible differences
in the magnetization data between $a$ and $b$ axis are also in
accord with recent report on untwinned crystals~\cite{hch}. Second,
at intermediate temperature, for all three field directions, a broad
maximum presents in the $m$ and $\chi_{m}$ curves, indicating the
short-range antiferromagnetic correlations of the low-dimensional
system. With the temperature increasing, a high-temperature
Curie-Weiss susceptibility is exhibited. In addition, it is worth
noted that the ratio of $J_{1}/J_{4}$ here is great than 1,
suggestive of a strong "rung" coupling in this compound.

To make a full comparison, in Fig. 6 we present the results of the
spontaneous polarization $\bm{P}$ and its electric susceptibility
$\chi_{e}$ along the $c$ axis, as well as the specific heat $C_{m}$
of the system. Similar to the experimental discovery, the two
magnetic phase transitions are also observed in our specific heat
data. For convenience and uniformity, the lower-temperature phase is
marked as $T_{N2}$ and the other is marked as $T_{N1}$. As shown in
Fig. 6(a), the main characteristics of $P_{c}$ observed in
LiCu$_{2}$O$_{2}$ are qualitatively reproduced based on the
spin-current model. Below the transition temperature $T_{N2}$,
$P_{c}$ gradually emerges and is accompanied by the sharp peak
arising in electric susceptibility $\chi_{e}^{c}$ (Fig. 6(b)) and
specific heat (Fig. 6(c)), and then rises rapidly and almost reaches
saturation below a lower temperature. When the system is polarized
by an opposite electric field, $P_{c}$ is reversed as well,
indicating the ferroelectric nature of spiral order
phase~\cite{ssy}. However, the reported remarkable field effects on
the electric polarization~\cite{sp} have not been exhibited in our
results. As shown in Figs. 6(a) and (b), $P_{c}$ and $\chi_{e}^{c}$
display the negligible responses for the field applied along
different directions. This discrepancy implies the profound effects
of the quantum fluctuations in LiCu$_{2}$O$_{2}$. In the specific
heat data, the phase transition at $T_{N1}$ is exhibited and
indicated by the smaller and round peak. Experimentally, it is found
that this transition is associated with collinear spin structure
with the spins sinusoidally modulated along the $c$ axis. Its origin
has not been explored yet. In the following, a conjecture and
corresponding analysis on the possibilities that induce this
transition will be given.

\subsection{The effects of various exchange interactions and exchange anisotropy}
In multiferroic system, spiral spin order is a common way to induce
the ferroelectricity and always yields some surprising physical
properties, such as flop, reversal and rotation of the electric
polarization in an external magnetic field. It usually generated by
the frustration in the magnetic materials, which origins from the
competition of various interactions. Therefore, analyzing the
impacts of different couplings on the spiral spin states is helpful
to understand the magnetoelectric properties in multiferroics.
\subsubsection{ "Rung" coupling $J_{1}$}
Figure 7(a) displays the influences of the "rung" coupling $J_{1}$
on the two transitions of the magnetic specific heat $C_{m}$. From a
general view, the two transition temperatures suffer the slight
affections and almost keep at the original positions, while their
corresponding peak values present the different responses. The
transition of $T_{N1}$ shows the ignorable variations, but the
transition of $T_{N2}$ exhibits the complex dependent relation on
$J_{1}$. For the sake of discussion, a parameter
$\alpha$=$J_{1}/J_{4}$ is introduced here. One can see that for
$\alpha\geq1$, an evident suppression on the peak of $T_{N2}$ is
presented. The ability of the system resisting the thermal
fluctuations is enhanced as well. While below $\alpha$=1, the
transition at $T_{N2}$ has no response to the variation of $J_{1}$.
These phenomenon are confirmed by the spin structure factors in
Figs. 7(b) and (c). At $\alpha$=1, the peak of $S(q)$ moves towards
large $q$ with its intensity decreasing, indicating the modulated
period of the incommensurate spin structure shortened by enhancing
$J_{1}$. These variations also demonstrate the fact that the spiral
order can be significantly modified under the strong rung coupling.
As $\alpha$ continues increasing, $S(q)$ shows very subtle decrease
on its peak value, suggestive of a balance of the energy competition
between the items of $J_{1}$ and $J_{4}$. However, it is worth noted
that a non-zero value appears at $q=0$ for $\alpha\geq1$, and it
acts in a more obvious way with the increase of $\alpha$, as shown
inset of Fig. 7(c). This unusual dependence implies the arise of the
low-temperature weak ferromagnetism.
\subsubsection{Next-nearest-neighbor inchain interactions $J_{4}$}
For a classical spin chain, when $|J_{4}/J_{2}|$ is larger than a
critical value 1/4, an incommensurate spiral spin structure with
pitching angle $2\pi\xi=\arccos(1/|4J_{4}/J_{2}|)$ can be expected
at ground state~\cite{sp}. And therefore the changes of $J_{4}$ will
evidently modify the magnitude of $\xi$ and are macroscopically
reflected in the variations of magnetic phase patterns. As shown in
Fig. 8(a), the two magnetic phase transitions both shift towards
high temperature with the increasement of $J_{4}$. The peak at
$T_{N2}$ transition becomes sharp and is greatly strengthened,
indicating the major adjustment of the spiral spin order, while the
one at $T_{N1}$ remains in a round shape and its almost constant
height. This is also vividly demonstrated in the picture of the spin
structure factors (Fig. 8(b)). When the value of $J_{4}$ is very
close to that of $J_{1}$, the peak of $S(q)$ is remarkably enhanced
and moves towards large $q$. As $J_{4}$ further increases, the peak
of $S(q)$ continues shifting to large value of $q$ with its
intensity invariant. This implies the reduction of the modulated
period and the improvement of spirality between nearest-neighbor
spins. In other words, it means the increasement of the pitch angle
and macroscopically enhancement of the polarization, which is also
well confirmed by our simulation results $P_{c}$ (not shown here).
In addition, at $q=0$, $S(q)$ starts at a small value and approaches
to zero finally, reflecting the furious competition between $J_{4}$
and $J_{1}$, which has been analyzed in Fig. 7.
\subsubsection{ exchange anisotropy $\Delta$}
Easy-plane anisotropy can enhance chiral correlation and is expected
to exist in the edge-sharing spin-chain compounds. For example, in
LiCuVO$_{4}$, a prominent of easy-plane anisotropy is probed by
ESR~\cite{hak}. This anisotropy is expected to pin the spins to the
$ab$ plane, which is also verified in experiment. In our simulation,
we find that it is also an indispensable factor for the occurrence
of two phase transitions. In Fig. 9, the temperature dependence of
$C_{m}$ for different anisotropies $\Delta$ are presented. Here the
"rung" coupling $J_{1}=2.4$ is made as the effects of the anisotropy
are more obvious in this way and the results are qualitatively
similar to those of $J_{1}=3.4$. As is shown, two humps appear with
the anisotropy introduced. With the increase of $\Delta$, one of the
two humps becomes sharp and moves to a higher temperature while the
other keeps almost at the original place.  At $\Delta=0.9$, the two
subpeaks merge into one, implying the existence of the
quasi-long-range spiral order and also indicated in the electric
susceptibility (not shown here). In fact, even though the easy-plane
anisotropy enhanced the spiral order, it suppresses the coupling
between spin along the $c$ axis, which is disadvantage for the
formation of the polarization. However, because of the large
magnitude of magnetic anisotropy $D$ chosen in this simulation, it
makes sure the strong spin coupling along the $c$ axis, stabilizing
the electric polarization arising at low temperature. Worth to
mentioned, we find that the easy-plane anisotropy is a critical
factor of inducing phase transition at $T_{N1}$, as it can generate
the competition of energies from the $ab$ plane spin coupling and
the $c$ axis spin coupling. Certainly, whether the spin
configuration is sinusoidal modulation at this phase transition is
still need further investigation. And we will discuss it in
somewhere else.

To analyze the variations of the microscopic magnetic and
ferroelectric structures, the $q$ dependence of $S(q)$ for different
$\Delta$ is calculated as shown in Fig. 10. At $\Delta$=0.4, the
maximum and sharp peak at $q$=0.32$\pi$ indicates the domination of
spiral order at $T_{N2}$. It worth noted that another two tiny peaks
also appear at $q$=0.02$\pi$ and $q$=$\pi$ respectively, which
suggestive of the existence of other possible magnetic phases at
ground state. Since our focus is on the spiral order in this paper,
the reason on the appearance of these two phases will not be
discussed here. With the increase of $\Delta$, the magnetic phase at
low temperature manifests as enhanced spiral order mixed with weak
ferromagnetism. As $\Delta\geq$0.7, only two characteristic sharp
peaks appearing at $q_{B}$ and its equivalent position
2$\pi$-$q_{B}$ signify the formation of the quasi-long-ranged spiral
order. To have a close view, the magnified of the collective
behaviors of $S(q)$ for various $\Delta$ are presented in Fig. 11.
As $\Delta$ increases, the tiny and round peak is enhanced and
become sharp with a shift towards small value of $q$, indicating the
increasement of the modulated period of spiral order. At
$\Delta$=0.7 and $\Delta$=0.8, the peak height keeps constant and in
the same position. A stable spiral order seems to be formed under
this circumstance. However, when $\Delta$ continues increase, the
peak of $S(q)$ with its invariant height shifts to large $q$,
demonstrating the complexity of $\Delta$ on the magnetic ordered
states.

\section{Conclusion}
In summary, based on the ellipsoidal spiral structure at ground
state, we qualitatively reproduce the complicated magnetoelectric
behaviors observed in LiCu$_{2}$O$_{2}$ by performing Monte Carlo
simulation. Our results indicate that the spin current model or the
inverse Dzyaloshiskii-Mariya mechanism still works even though large
quantum fluctuation exists in this compound. The spin structure
factors have been evaluated to confirm the spiral spin order at
lowest temperature and detect the variations of the microscopic
magnetic structure. The divers effects of the different exchange
couplings and exchange anisotropies on the magnetic patterns have
been vividly demonstrated in this simulation. It turns out that the
"rung" coupling $J_{1}$ can not be simply neglected, and the
exchange anisotropy has a complicated influences on the magnetic
order. Besides, the deviation of the field responses of the
polarization from the experiment implies the importance of quantum
fluctuation in this S=1/2 spiral magnet, which requires further
studies on this interesting ferroelectric cuprate.\\

\section{Acknowledgements}
 This work was supported by Academic Scholarship for Doctoral
Candidates of China (Grant No. 10145201103), the Fundamental
Research Funds for the Central Universities of China (Grant No.
N110605002) and Shenyang Applied Basic Research Foundation (Grant
No. F12-277-1-78) of China.

\newpage
\begin{figure}
\begin{center}
\includegraphics[scale=0.5]{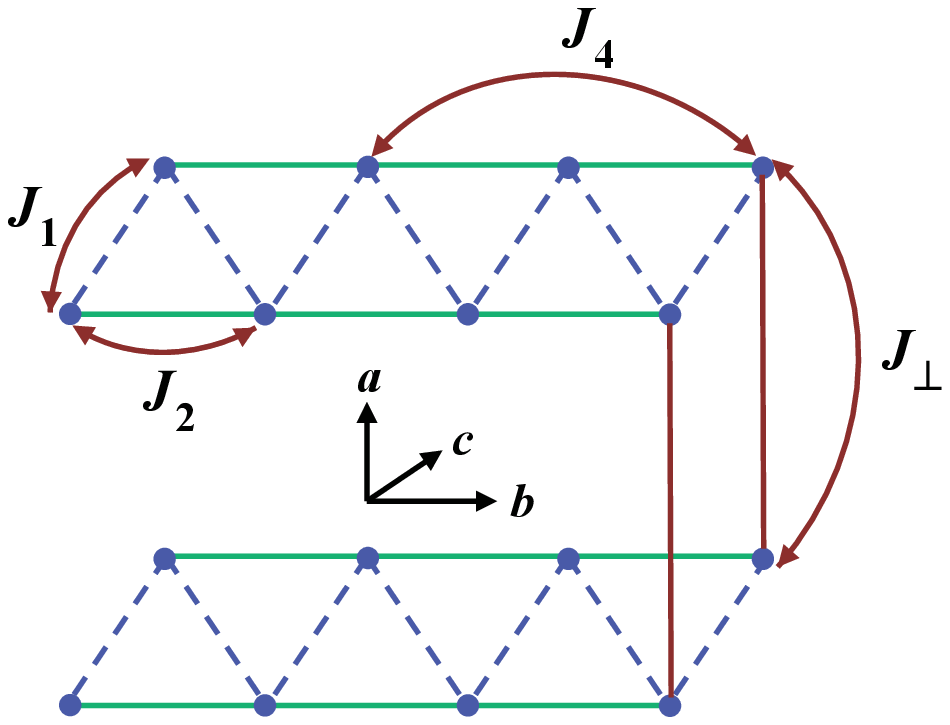}
\end{center}
\caption{(Color online) A schematic view of exchange interactions
between magnetic Cu$^{2+}$ ions in LiCu$_{2}$O$_{2}$.}
\end{figure}

\begin{figure}
\begin{center}
\includegraphics[scale=0.3]{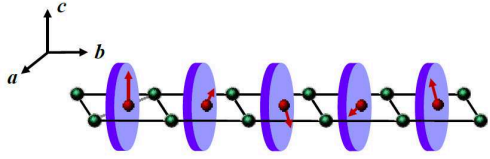}
\end{center}
\caption{(Color online) Schematic view of the spin configuration of
LiCu$_{2}$O$_{2}$ in the ground state. The green and red balls
represent O$^{2-}$ and Cu$^{2+}$ ions, respectively.  }
\end{figure}


\begin{figure}[H]
\centering \subfigure{
\includegraphics[width=0.35\textwidth]{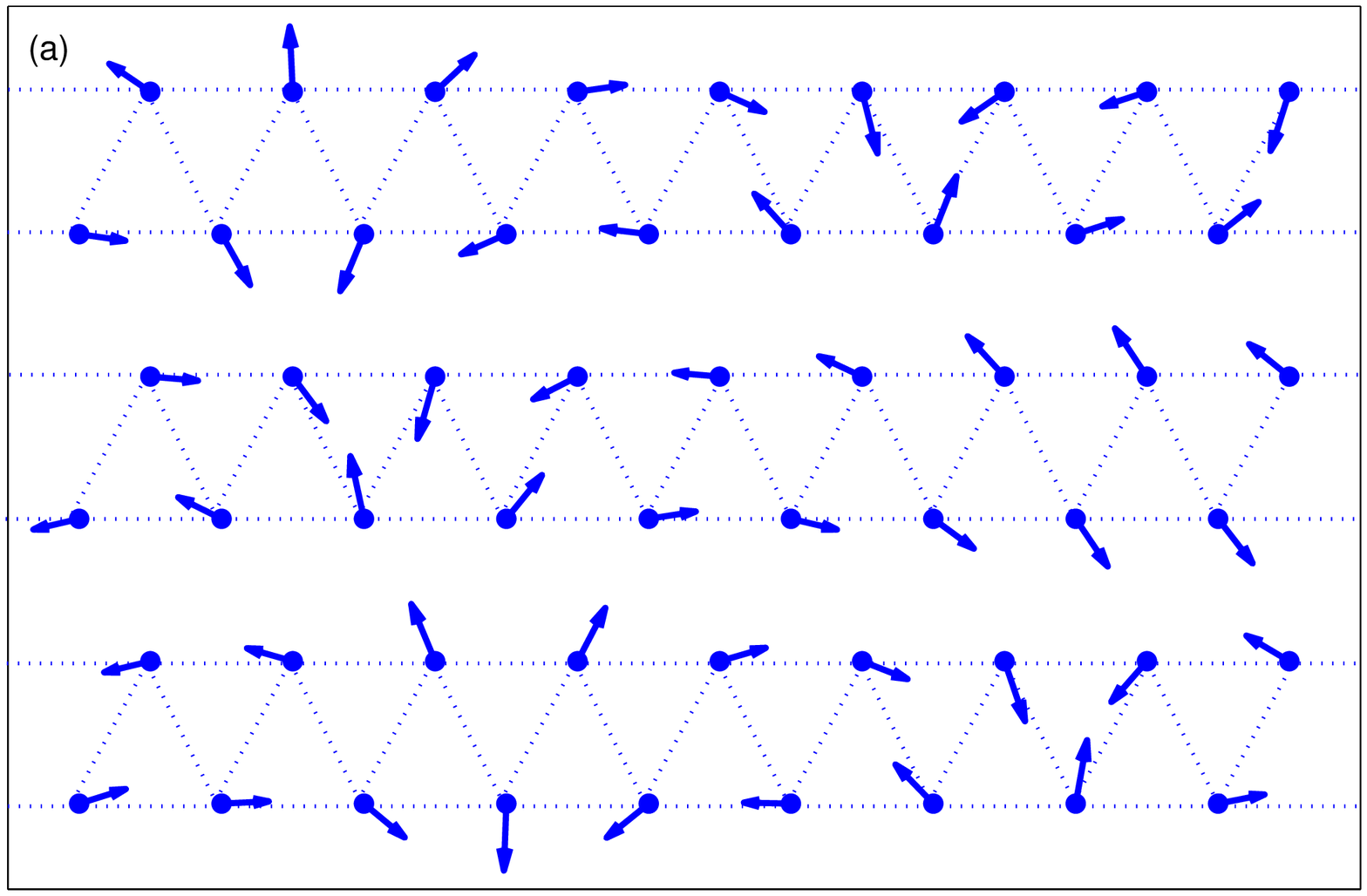}}
\subfigure{
\includegraphics[width=0.35\textwidth]{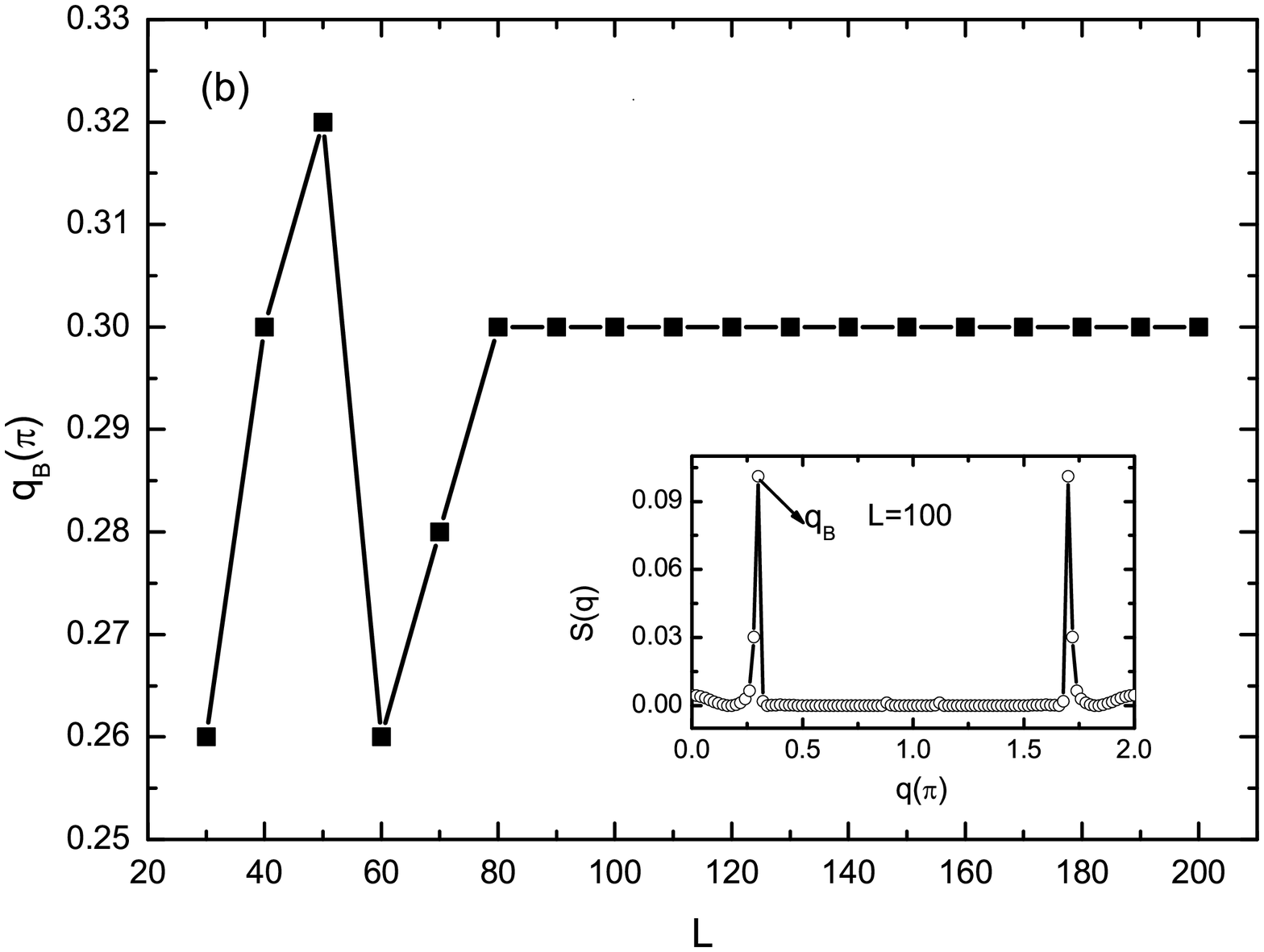}}
\caption{(Color online) (a) A projection
 of the spin configuration
to the $bc$ plane at $T$=0.01. (b) The lattice size dependence of
wave vector $q_{B}$ at $T$=0.01. The inset is spin structure factor
$S(q)$ for $L=100$.} \label{Fig.lable}
\end{figure}

\begin{figure}
\begin{center}
\includegraphics[scale=0.2]{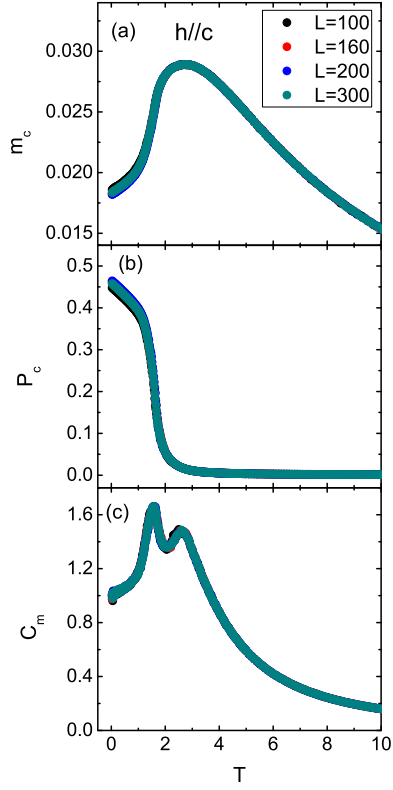}
\end{center}
\caption{(Color online) Temperature dependence of the magnetization,
polarization and magnetic specific heat for different lattice sizes
$L$ with $h=0.5$ applied along $c$ axis.}
\end{figure}

\begin{figure}
\begin{center}
\includegraphics[scale=0.4]{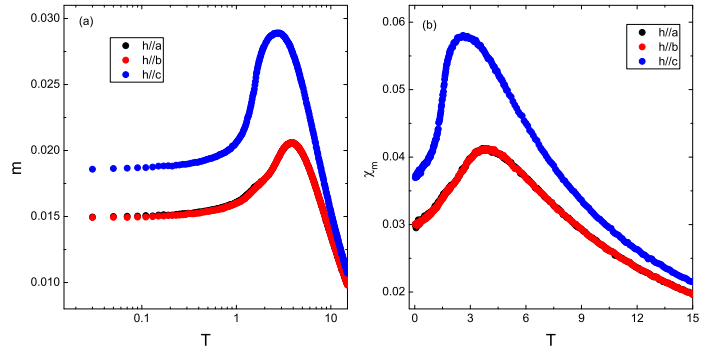}
\end{center}
\caption{(Color online) Temperature dependence of the magnetization
curves (a) and susceptibility curves (b) under $h=0.5$ and applied
along three axis respectively. }
\end{figure}

\begin{figure}
\begin{center}
\includegraphics[scale=0.2]{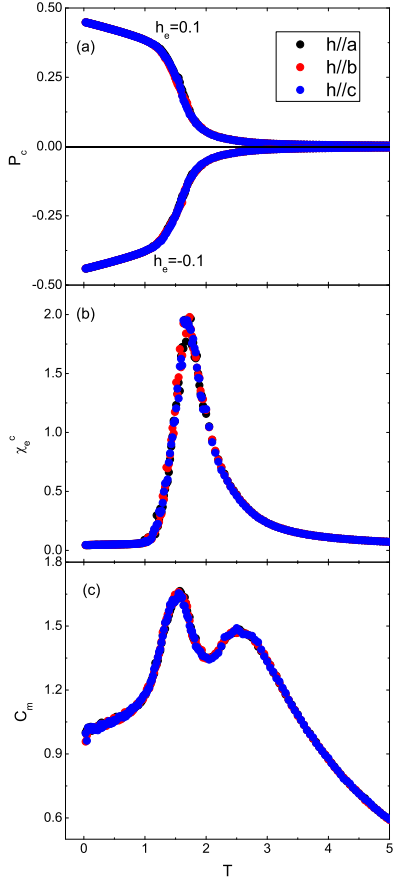}
\end{center}
\caption{(Color online) Temperature dependence of (a) the
polarization $P_{c}$ along the $c$ axis, (b) the corresponding
electric susceptibility $\chi_{e}^{c}$ and (c) the magnetic specific
heat $C_{m}$ for three field directions under $h$=0.5.}
\end{figure}

\begin{figure}[H]
\centering \subfigure{
\includegraphics[width=0.35\textwidth]{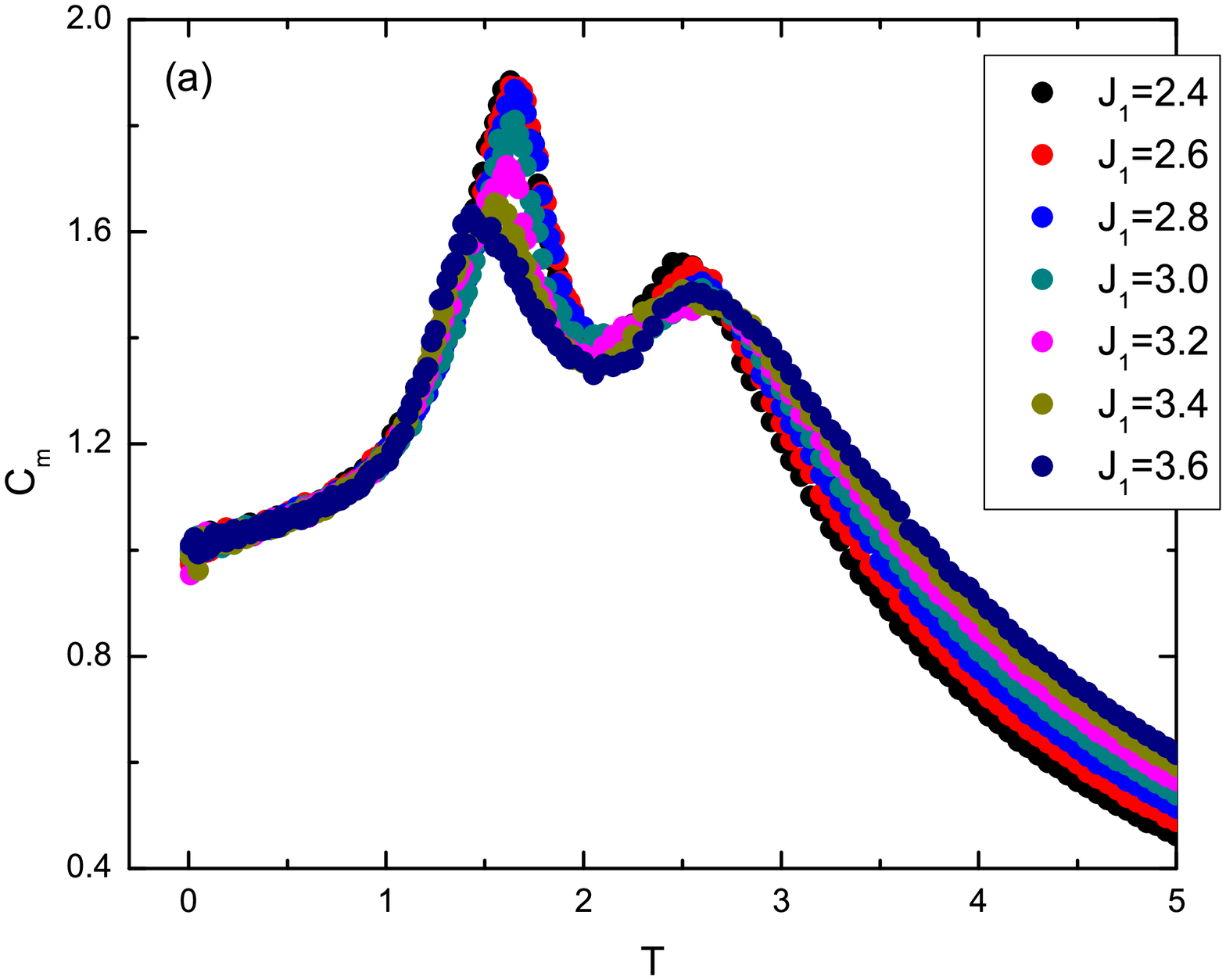}}
\subfigure{
\includegraphics[width=0.6\textwidth]{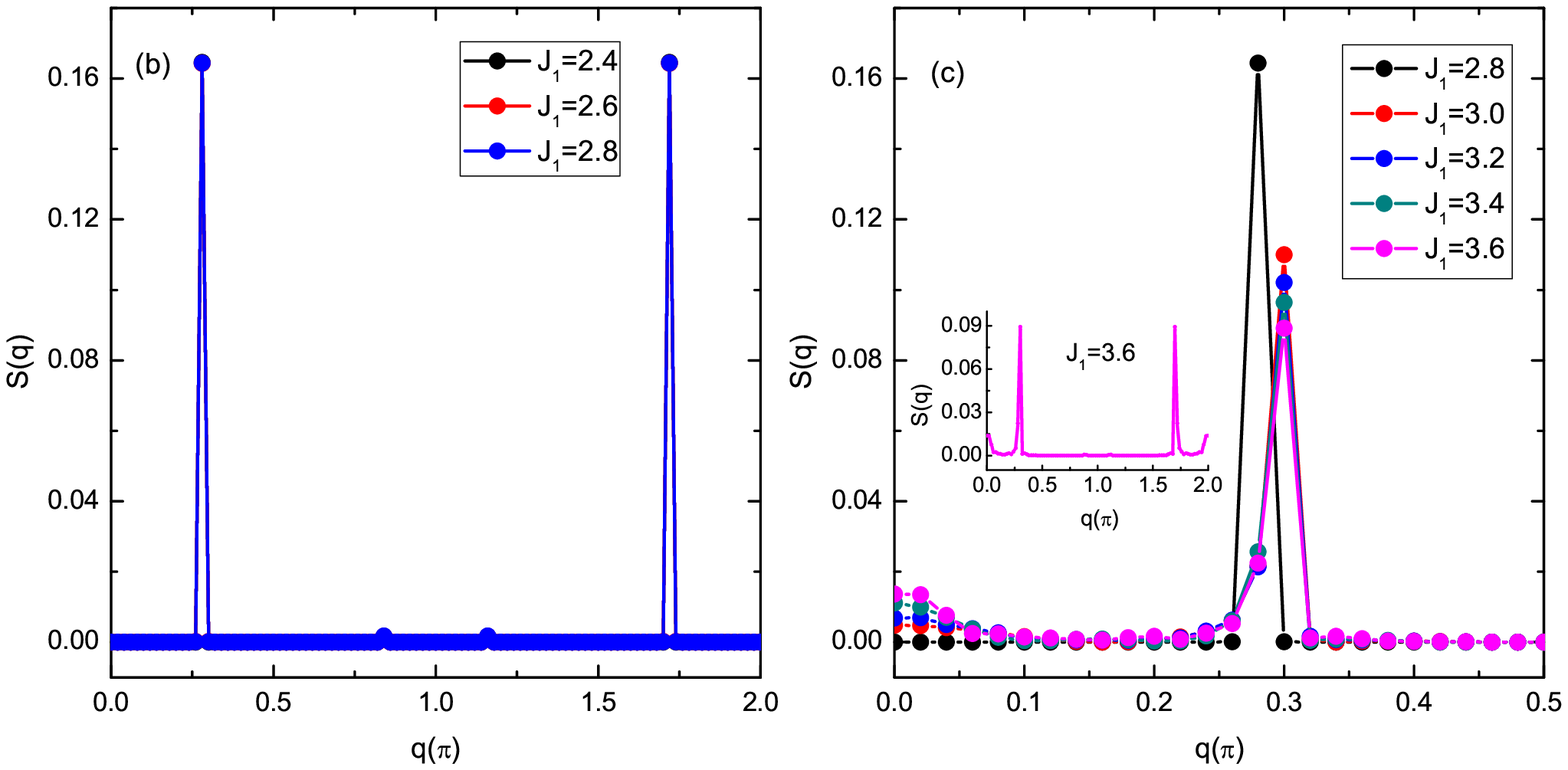}}
\caption{(Color online) (a) Temperature dependence of the magnetic
specific heat with the field $h=0.5$ applied along the $c$ axis for
different "rung" coupling $J_{1}$. (b), (c) Spin structure factors
for different "rung" couplings at $T=0.01$. The inset is the spin
structure factor $S(q)$ for $J_{1}=3.6$.} \label{Fig.lable}
\end{figure}

\begin{figure}
\begin{center}
\includegraphics[scale=0.4]{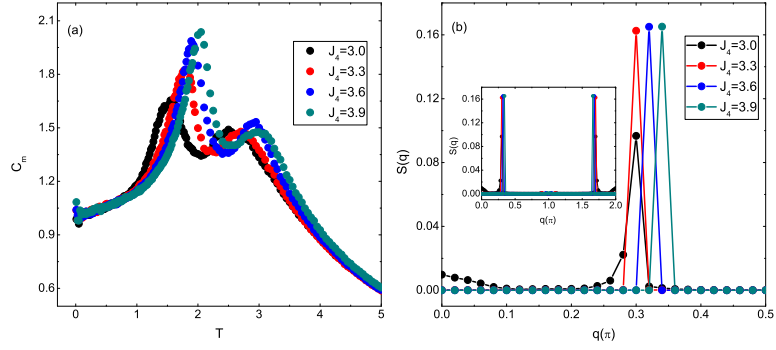}
\end{center}
\caption{(Color online) (a) Temperature dependence of the magnetic
specific heat with the field $h=0.5$ applied along the $c$ axis and
(b) spin structure factors $S(q)$ at $T=0.01$ for different
next-nearest neighbor inchain coupling $J_{4}$. The inset is the
full view of spin structure factors $S(q)$. }
\end{figure}


\begin{figure}
\begin{center}
\includegraphics[scale=0.3]{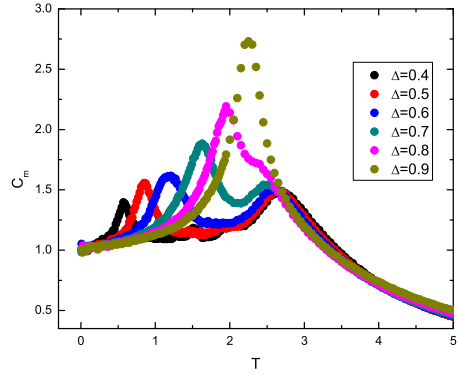}
\end{center}
\caption{(Color online) Temperature dependence of the magnetic
specific heat with the field $h=0.5$ applied along the $c$ axis for
different anisotropic coupling $\Delta$.  }
\end{figure}

\begin{figure}
\begin{center}
\includegraphics[scale=0.4]{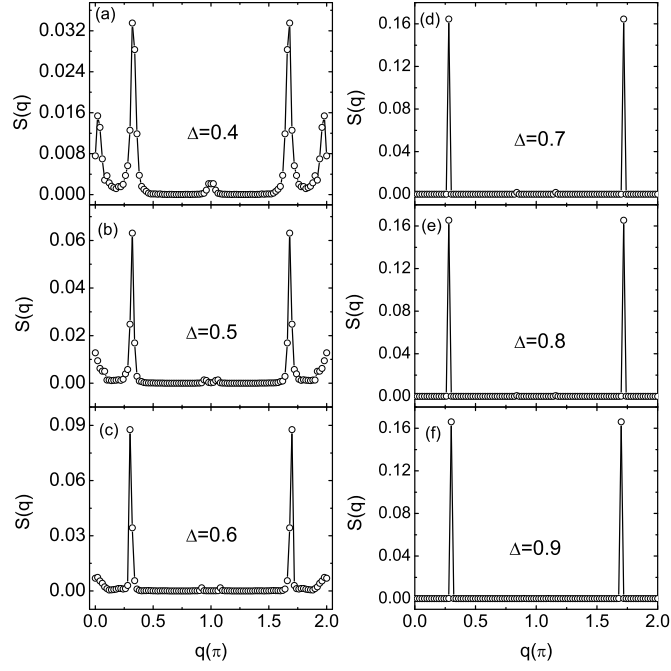}
\end{center}
\caption{(Color online) (a)-(f) The wave vector dependence of spin
structure factor $S(q)$ for different exchange anisotropic couplings
$\Delta$ at $T=0.01$. }
\end{figure}

\begin{figure}
\begin{center}
\includegraphics[scale=0.3]{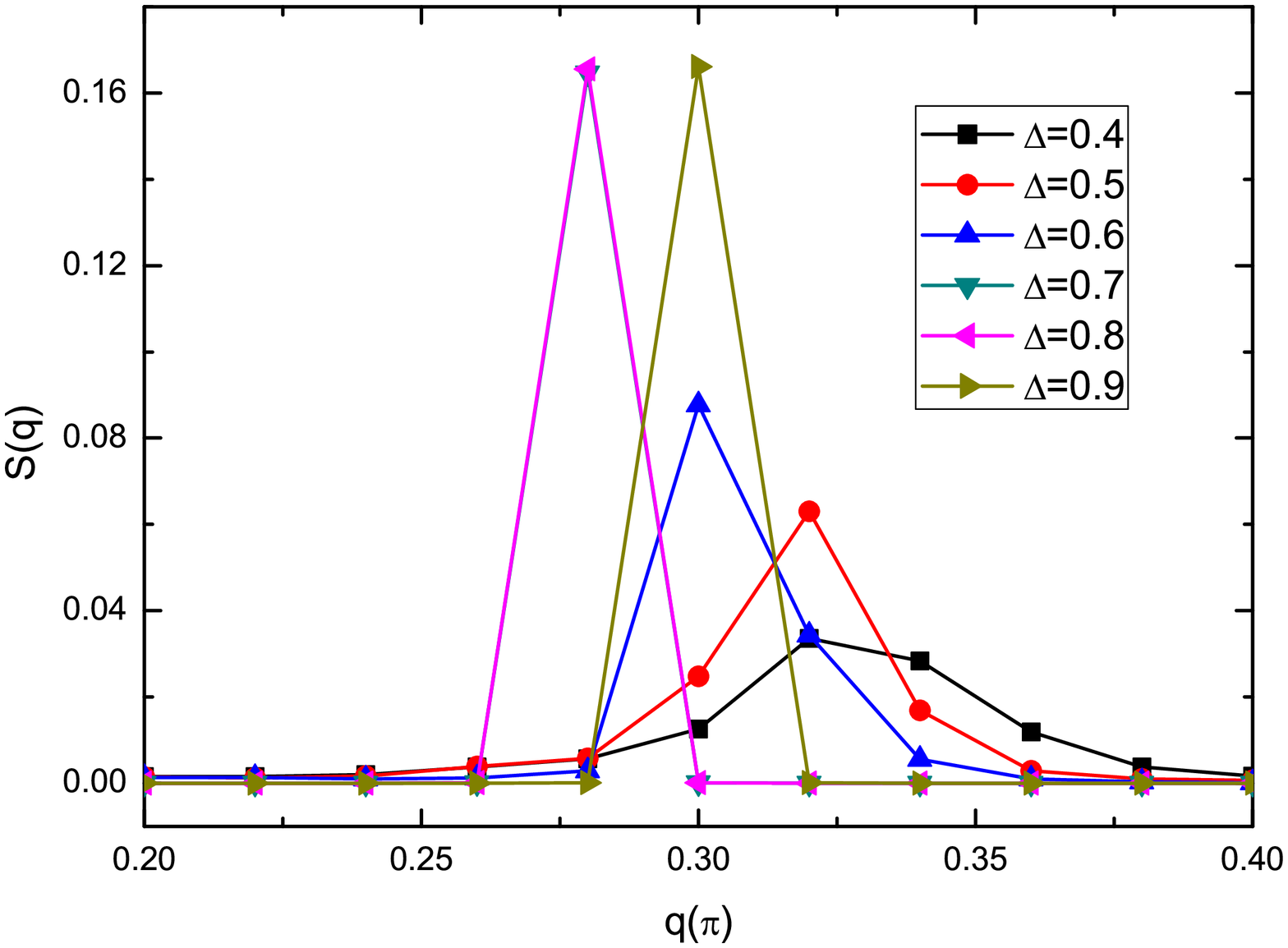}
\end{center}
\caption{(Color online) The enlarged view of spin structure factors
$S(q)$ under various exchange anisotropies $\Delta$. }
\end{figure}
\end{document}